\documentstyle[preprint,aps]{revtex}
\begin{document}
\draft
\widetext
\title{ Large physical spin approach for strongly correlated
electrons }
\author { Antimo Angelucci }
\address {Institut f\"ur Theoretische Physik, Universit\"at W\"urzburg,
D--97074 W\"urzburg, Germany}
\author{Sandro Sorella}
\address {International School for Advanced Studies,
Via Beirut 2-4, I--34013 Trieste, Italy}
\author { Didier Poilblanc }
\address { Laboratoire de Physique Quantique,
Universit\'e  Paul Sabatier, F--31062 Toulouse, France}
%
%
%
%

\maketitle
\begin{abstract}
We present a novel approach for a systematic large--spin expansion of the
$t$-$J$ Hamiltonian which enables us to work without the constraint of no
double occupancy. In our scheme we can perform the large--spin limit ensuring
that the low energy spin excitations are  in {\em exact}  correspondence with
the physical excitations of the $s={1\over2}$ Hilbert space. As a consequence,
we expect a smooth dependence of the physical quantities on the expansion
parameter $1/ s$. As a first application of the method we study the case of
a single hole in a N\'eel background. A systematic
expansion in fluctuations about this stable solution indicates that
by increasing $t/J$ the quasiparticle weight strongly depends on the
momentum carried by the hole.  Results, obtained on small lattice sizes,
 are found in excellent agreement with exact diagonalization data.
\end{abstract}

\narrowtext

The $t$-$J$ model in two spatial dimensions is perhaps the most challenging
``unsolved'' problem in the theory of strongly correlated electrons,
since it is now commonly accepted to represent the low--energy
Hamiltonian for the two dimensional copper--oxide high--temperature
superconductors. Recent calculations based on
the old fashioned, but reliable high--temperature
expansion techniques, have indicated that spin--charge separation,
obviously present in this model  in one dimension, may also
characterize the elementary excitations in 2D, leading to a
break-down of Fermi liquid theory and to a possible explanation of the
anomalous properties of the high--temperature superconductors.\cite{singh}

We consider $N_h$ holes interacting by the $t$-$J$ Hamiltonian
$$
H_{tJ} =  -t \sum \limits_{\langle i,j\rangle,\sigma } P
(c^{\dag}_{i,\sigma} c_{j,\sigma} + h.c. ) P  +
          J \sum \limits_{\langle i,j\rangle}
(\vec S_i \, \vec S_j - {1\over 4 } N_i \, N_j),
$$
where $\langle i,j\rangle$ denotes a summation over the nearest neighbor
(n.-n.) sites
of the lattice, $P$ is the projector onto the Hilbert space without doubly
occupied sites, $\vec S_i $ and $N_i$ are the spin and number
operators at site $i$, and $c^{\dag}_{i,\sigma}$ and $c_{i,\sigma}$ are the
usual creation and  annihilation operators for electrons of spin $\sigma$.
Henceforth we assume $t\ge 0$, $J\ge 0$.

Due to the difficulty to deal with the projector of no double occupancy,
several semiclassical approaches\cite{affleck,kotliar,klr,auerbach}
leading to a mean--field description of $H_{tJ}$ have been proposed.
In particular, in the large--spin approaches presented so far, the
simplification of the $t$-$J$ Hamiltonian is achieved by generalizing
the spin-${1\over 2}$ polarization state of the electron to an arbitrary
spin $s$ state and by giving a fictitious spin $(s-{1\over 2})$ to the hole.
The various methods differ for the definition of the enlarged Hamiltonian, a
freedom left by the very fact that in a large--spin
generalization one can only require that the physical Hilbert space, as well
as the $t$-$J$ model $H_{tJ}$, must be recovered for the value
$s={1\over2}$ of the
expansion parameter. Unfortunately, all of them
face the following fundamental difficulty: {\em
as soon as the extended Hilbert space is larger than the physical one,
spurious low--energy elementary excitations  emerge}.
Hence, it is not at all guaranteed that the low--lying
excitations of the extended Hamiltonian correspond to some physical
excitation of the original one, so that it becomes difficult or even
impossible to derive reliable results by performing a systematic expansion
in fluctuations about the mean--field obtained by letting $s \to \infty$.

Explicitly, in the
Kane {\it et al.} model one obtains the N\'eel background as the
mean--field solution for small $J/ t$, but for $s>{1\over2}$ some of the
allowed excitations change the spin of the hole, which is clearly
unphysical. Other large--spin generalizations instead face the problem that
the hole propagation becomes a nonperturbative process
in the $1/s$ expansion. In the latter
case, as was shown by two of us in Ref.\cite{antimo},
one obtains phase separation even at small $J/t$, whereas
it is now believed  that the uniform ground--state is stable in the physical
sector \cite{putikka,zhong}.


In this work we propose a new method to simplify at large spin the $t$-$J$
model without facing the previous difficulties. In our approach we do not
deal directly with $H_{tJ}$, but consider instead a natural extension
of it as introduced by Sutherland\cite{suth}, because the latter
allows us to apply the spin--wave theory with a one--to--one correspondence
with the physical excitations.


In the $t$-$J$ Hamiltonian the single site $i$ can be occupied by 3
kinds of ``objects'': A hole (boson) $\vert 0 \rangle_i$, an electron of
spin-up
$\vert \!\uparrow\, \rangle_i=c^{\dagger}_{i,\uparrow}\vert 0 \rangle_i$,
and an electron of spin-down
$\vert \!\downarrow\, \rangle_i=c^{\dagger}_{i,\downarrow}\vert 0 \rangle_i$,
(fermions) whereas, apart for an irrelevant energy shift, $H_{tJ}$
can be thought of as
the operator permuting pairs of n.-n. objects,
with weight $t$ for permutations of objects of opposite statistics, and
weight $J/2$ ($-J/2$) for permutations of fermions (bosons).
In order to work without the local constraint of no doubly occupancy, we
consider the extended Hamiltonian $H$ acting on objects of two fermion and
{\it two} boson species by permutation of pairs of neighboring objects
with the same weights as in the $t$-$J$ model. Because the number of objects
of a given species is conserved by construction, the reduction to the
physical model can be obtained
by projecting onto the invariant subspace where one boson species is absent.
Hence, in our approach the projector operator $P$ is washed out, as for
the projection amounts just to fix a conserved quantity.

To represent the extended Hamiltonian $H$ in a way suited for our
developments, we denote the fermion and boson objects
at site $i$ with the symbols $\vert 1\sigma\rangle$ and
$\vert 0\sigma\rangle$ ($\sigma=\uparrow,\downarrow$), respectively,
and use the representation
$ \vert 0\!\uparrow\, \rangle_i = f_i^{\dagger}   \vert v \rangle$,
$ \vert 1\!\uparrow\,  \rangle_i =                \vert v \rangle$,
$ \vert 0\!\downarrow\, \rangle_i = f_i^{\dagger}Q_{i,-} \vert v \rangle$,
$ \vert 1\!\downarrow\, \rangle_i =              Q_{i,-} \vert v \rangle$,
where $f_i^{\dagger}$ and $Q_{i,-}$ are a spinless
fermion creation operator and a spin-${1\over 2}$ lowering operator,
respectively.
The extended Hamiltonian reads
$$
H = \sum \limits_{\langle i,j \rangle} \left[
-t\,(f_i f^{\dag}_j + f_j f^{\dag}_i)\chi_{i,j}   +
{J \over 2} (1-n_i -n_j) (\chi_{i,j}-1) \right],
\nonumber
$$
where $\chi_{i,j} = 2 \vec Q_i \, \vec Q_j \,+ \, {1\over 2} $ and
$n_i=f_i^{\dagger}f_i$. A noticeable feature of the proposed model $H$, to be
contrasted with the case of $H_{tJ}$ in a slave--fermion
representation\cite{auerbach,antimo}, is the presence of the permutator
operator $\chi_{i,j}$ both in the magnetic and {\it kinetic} part, as well
as the bilinear dependence on the fermion operators.

Henceforth we preserve the name of electron and hole for the
two ``particles'' $\vert 1\sigma\rangle$ and $\vert 0\sigma\rangle$,
respectively, and introduce the two commuting
vector operators satisfying the algebra of the angular momentum
\begin{equation}
\vec S \,=\, \sum\limits_i (1-n_i) \vec Q_i , \qquad
\vec L \,=\, \sum \limits_i n_i \vec Q_i ,
\label{spin}
\end{equation}
The operators $\vec S$ and $\vec L$ act  nontrivially on
$\vert 1\sigma\rangle$ and $\vert 0\sigma\rangle$, respectively, and
accordingly $\vec S$ will be referred to as the physical spin and $\vec L$
as the pseudospin.
The analogy with the properties of the spin and pseudospin operators is not
only formal and will be discussed elsewhere\cite{papnew}. In the following
we shall also refer to $\vec Q = \vec L + \vec S$ as the isospin vector.
The operators (\ref{spin}) commute with  $H$, so that the
quantum numbers spin $S_z$, total spin $S$, and pseudospin $L_z$,
total pseudospin $L$ associated to both ``particles'' are conserved.
The physical Hilbert space of $H_{tJ}$ corresponds to the sector
where the pseudospin attains its maximum value  $L_z = L = N_h/2$.

Sutherland\cite{suth}  has shown  quite generally  that {\em the
ground--state of the Hamiltonian} ($H$) {\em is at most degenerate
with the physical one with  maximum pseudospin} $L_z$. Unfortunately,
this statement is rigorously valid only in one dimension, and for the
case of a single hole in any dimension. The latter case is of course
trivial, because for one hole the pseudospin is by definition equal to
the maximum value $L={1\over2}$. The proof presented in Ref.\cite{suth}
is not valid in 2D. In fact, following the reasoning one would obtain that
for $J=0$ the ground--state of $H_{tJ}$ is the fully polarized Nagaoka
state, whereas it is known that for large doping the singlet Gutzwiller
projected Fermi gas has macroscopically lower energy\cite{shiba}.

However, the Sutherland's result is true for the special case
$t=0$ and probably remains
valid for physically acceptable $J/t>0$ in D$\ge$2. Hence, we shall leave
it as a ``conjecture''. The importance of this conjecture is easily
understood by noting that whenever it is satisfied, one can evaluate
ground--state properties avoiding even the projection onto the $L_z=N_h/2$
sector.

The Hamiltonian $H$ still represents a highly nontrivial problem
and we now consider the large--spin approach allowing to simplify the model.
Noting that $\vec Q$ is a irreducible spin-${1\over2}$ operator,
we consider arbitrary higher--dimensional representations of the isospin
vector and define the enlarged Hamiltonian $H_s$ by substituting in the
extended Hamiltonian $H$  the permutator operator $\chi_{i,j}$ with
the rotationally invariant expression
\begin{equation}
\chi_{i,j} \to
\chi^{(s)}_{i,j} = {1 \over 2  s^2 } \vec Q_i \,\vec Q_j +  {1\over 2} .
\label{chi}
\end{equation}
The overall factors and constants -- irrelevant in the undoped case --
are set by the requirement that $\langle \chi^{(s)}_{i,j}\rangle$
is one or zero if the isospins of the particles at sites $i$,$j$ are
parallel or antiparallel, respectively. In our approach
the Hilbert space is generalized by giving a fictitious spin-$s$
both to the electron (i.e., $\vert 1\sigma\rangle$) and to the
{\it hole} (i.e., $\vert 0\sigma\rangle$) and in this respect it is quite
different from the large--spin approaches proposed so far.

%

Because at zero doping isospin and physical spin coincide ($\vec L=0$),
our approach leads to the conventional spin wave--expansion
for the Heisenberg antiferromagnet: The ground--state
is a singlet and the physical spin--wave excitations have $S=1$,
i.e., they are independent of the magnitude of the spin $s$ of the
extended Hilbert space. We see that in the
large--spin limit the Hilbert space at each single site grows with $s$ but
the low--energy excitations remain in one--to--one correspondence with those
of the physical $s={1\over2}$ Hilbert space. We believe that this is the
basic reason why the spin--wave expansion is so accurate for the undoped
system and why $1/s$ is a smooth parameter and actually
small\cite{zhong,canali,runge,young}.

At nonzero doping, for $s>{1\over2}$
the observable (\ref{spin}) are no more conserved, unless for $t=0$.
However the Hamiltonian $H_s$ always commutes with the isospin
and it is therefore convenient to work in the subspace with $Q$ and $Q_z$
fixed. If for the original $t$-$J$ Hamiltonian the physical spin
attains its minimum value - as it is expected for $J$ not too small --
the Hilbert space with minimum $S$ and fixed $S_z$ is exactly
equivalent to the one
with $Q=[N_h/2]$ and say $Q_z=0$ (for $N_h$ odd one has
$S=1/2$ as the minimum value). Hence, if we are able to classify the
elementary excitations for $s\to\infty$ in the sector where $L$
is frozen to its maximum value, the same classification would hold
in the physical Hilbert space, provided that there is no phase transition
as a function of $1/s$.

In order to show that our $1/s$ expansion is the natural extension of
the spin--wave theory even in the doped case, we focus our attention
to the simplest possible nontrivial doping, i.e., when there is only one
hole ($N_h=1$)  in  a lattice of $M$ sites  with periodic boundary
conditions.

For large $s$ and $J$ not too small the stable semiclassical
solution corresponds to a N\'eel background where the single hole can
propagate with given momentum either on the $A$ or on the $B$ sublattice.
Fluctuations over this semiclassical solution are obtained in the
usual way by introducing boson operators
$a^{\dag}_i \approx {  Q_{i,-} \over  \sqrt{2s} } \,$
if $i\in A$ (${ Q_{i,+} \over \sqrt{2s}} \,$ if $i\in B$) that create a
spin fluctuation over the N\'eel classical state
$\vert N \rangle$. Thus a systematic
expansion of the operator $\chi^{(s)}_{ij}$ in $1/s$ is possible and
we get
\begin{equation}
\chi^{(s)}_{i,j}= {1\over 2 s} ( \psi_{i,j}^{\dag} \psi_{i,j} -1) \,+\,
O({1\over s^2}) ,
\label{chiexp}
\end{equation}
where $\psi_{i,j}=a^{\dag}_i + a_j$. By replacing the expression
(\ref{chiexp}) in the Hamiltonian $H$, both in the kinetic and
the magnetic term, we then obtain an effective Hamiltonian for
the single hole, which is characterized  by a  kinetic term coupling
two boson and two fermion operators
$ {-t\over 2 s} ( f_i f_j^{\dag} +\,{\rm h.c.} )[ (a^{\dag}_i+a_j) (a_i +
a^{\dag}_j) -1  ]$. Hence, it is remarkably different from the Kane
{\it et al}. Hamiltonian, where instead the hole propagates
by emitting or absorbing a single spin fluctuation. In our approach
the spin is carried only by the boson $a^{\dagger}_i$
(which changes the spin by one), so that
the conservation of $Q_z$  necessarily implies a quadratic
Hamiltonian in the boson operators. A further simplification of $H_s$
can be obtained following Ref.\cite{sorella,edwards}
for a different but similar problem.
One can exactly trace out the single fermion $f_i^{\dag}$ from the
Hamiltonian $H$ using translation invariance and thus obtaining an effective
spin Hamiltonian  defined by the  translation operator
$T_{\tau_\mu} a_i T_{-\tau_\mu}=a_{i+\tau_\mu}$ of spin--waves for
nearest neighbour displacements $\tau_{\mu}$
\begin{equation}
H_{\rm eff}\,=\, {1 \over 4 s} \sum\limits_{\tau_{\mu}}
(\psi_{0,\tau_{\mu}}^{\dag} \psi_{0,\tau_{\mu}} -1)  ( 2 t\,
T_{\tau_{\mu}} e^{ i p \tau_{\mu}} - J ) + H_{SW},
\label{tjhole}
\end{equation}
where $i=0$ denotes the origin and $H_{SW}$ is the Heisenberg
Hamiltonian that, at first order in $1/s$ reads:
$H_{SW}= {J \over 4 s} \sum\limits_{\langle i,j \rangle}
(\psi_{i,j}^{\dag} \psi_{i,j} -1)$.

Contrary to the undoped case, the Hamiltonian $H_{\rm eff}$
cannot be solved analytically unless for the case $t=0$,
 where  $H_{\rm eff}$ becomes
quadratic\cite{mahan,sorella}.
However, a very good variational wavefunction that is exact in this
limit, and which preserves all the symmetries of the Hamiltonian, is very
easy to write down, in the form of the most general ground--state of a
quadratic Bogoliubov Hamiltonian:
\begin{equation}
\vert \Psi_h \rangle \,=\,
\exp\{{1\over 2} \sum\limits_{i,j} B_{i,j}
a^{\dag}_i a^{\dag}_j \}  \vert N \rangle .
\label{psig}
\end{equation}
$B_{i,j}$ is non zero only if $i$ and $j$ belong to different
sublattices (to fulfil $Q_z=0$) and its Fourier transform does not
contain the modes at $k=(0,0)$ and $k=(\pi,\pi)$  (to fulfil $Q=0$)
\cite{zhong}. In order to evaluate and then minimize the expectation value
of the Hamiltonian $H_{\rm eff}$ over the state (\ref{psig}),
one needs to evaluate both the average
$\langle  \Psi_h\vert  \psi_{i,j}^{\dag} \psi_{i,j} \vert
\Psi_h \rangle$,
and the average of the quadratic form $\psi_{i,j}^{\dag} \psi_{i,j}$
over $\vert \Psi_h \rangle$ and the state
$\vert \Psi_h^\mu \rangle =T_{\tau_\mu} \vert \Psi_h \rangle$ generated
by the translation operator entering the kinetic term,
i.e., $\langle  \Psi_h\vert  \psi_{i,j}^{\dag} \psi_{i,j} \vert
\Psi_h^\mu  \rangle $.
The state $\vert \Psi_h^\mu  \rangle$ is clearly
obtained by replacing in Eq.\ (\ref{psig}) the matrix $B_{i,j}$ with
$B^{\mu}_{i,j}=B_{i-\tau_\mu,j-\tau_\mu}$ and
because both the states $\vert \Psi_h^\mu\rangle $ and
$\vert \Psi_h\rangle$ are gaussian, it is then possible
to evaluate the averages in closed form. In fact, given two gaussian
states  $\vert \Psi_A \rangle $ and $\vert \Psi_C \rangle$ of the form
(\ref{psig}), with $A$ and $C$ the corresponding matrices entering the
exponential, we have
\begin{eqnarray*}
{ \langle \Psi_A \vert a_i a^{\dag}_j \vert \Psi_C \rangle \over
\langle \Psi_A \vert  \Psi_C \rangle } &=& G_{i,j}
 \\
{ \langle \Psi_A \vert a^{\dag}_i a^{\dag}_j \vert \Psi_C \rangle \over
\langle \Psi_A \vert  \Psi_C \rangle } &=& \left[
A^* G  \right]_{i,j} \\
{ \langle \Psi_A \vert a_i a_j \vert \Psi_C \rangle \over
\langle \Psi_A \vert  \Psi_C \rangle } &=& \left[
G  C \right]_{i,j} \\
\end{eqnarray*}
where
\begin{eqnarray}
G_{i,j}&=& { \langle \Psi_A \vert \Psi_C \rangle \over
\langle \Psi_A \vert \Psi_A \rangle }
\left[(I -  C A^* )^{-1}\right]_{i,j} \nonumber \\
\langle \Psi_A \vert \Psi_C \rangle &=& {\rm det}^{-1/2}
  (I -  C A^* ) \label{det}
\end{eqnarray}
Using the above equations is it easy to work out an explicit expression
for the expectation value of the Hamiltonian by simple linear algebra
operations over the symmetric complex matrix $B_{i,j}$.
We have then obtained the optimal matrix $B_{i,j}$
by minimizing the energy  of the effective Hamiltonian  $H_{\rm eff}$ with
the standard conjugate gradient technique.

In table I we show  data for the one hole energy  (referenced
to the undoped energy) and the quasiparticle weight
$Z=\vert \langle \Psi_H \vert \Psi_h\rangle \vert ^2$,
where $\vert \Psi_H \rangle$ is  the ground--state of $H_{SW}$.
Because $\vert \Psi_H \rangle$ is easily written in the gaussian form
(\ref{psig})\cite{klr,zhong}, Eq.\ (\ref{det}) allows to evaluate
$Z$ straightforwardly.
The agreement of the spin--wave estimates with the exact diagonalization
results\cite{didier} is surprisingly good, yielding a robust evidence of a
finite value for $Z$ in the static $t=0$ limit.

The accuracy of the method remains very good even for $t>0$,
as it is shown in Fig.1 for the quasiparticle weight at $p=(0,0)$
and $p=(\pi,\pi)$. We also see in Fig.1 (b) a clear transition of the
quasiparticle weight for the $(\pi,\pi)$ momentum. Its value  changes of about
two order of magnitudes also in the exact diagonalization data.

Our spin--wave approximation agrees with the exact diagonalization even
in the details for $t/J<1$.
In fact, at the value $t/J \simeq 0.5$,  where in our
simulation we find  a singular point [see Fig.1 (b)], there is a true
level crossing in the exact diagonalization. The true ground--state is
actually orthogonal (with different symmetry) to $\vert \Psi_H \rangle$.
For larger $t/J$ our approximate solution predicts that  $Z_{(\pi,\pi)}$
vanishes at  a critical point for any finite size $M$.
At the moment a similar analysis\cite{newsandro} for momenta close to
$({\pi\over2},{\pi\over2})$
(which is found to be always the ground--state for $M \to \infty$,
consistent with the general believe in the physical region $J$ not
too small)
indicates that $Z_{({\pi\over2},{\pi\over2})}$ remains finite
up to $t/J \simeq 3 $\cite{sorella}.
However we cannot exclude a transition to $Z_{({\pi\over2},{\pi\over2})}=0$
at larger values of $t/J$ where our variational approach becomes
unreliable.

The fact that some of the low energy excitations of the Mott insulator
may have a ``non trivial'' character may lead to a
completely new classification  of the charge excitations in the low
doping regime.
This surely requires further analysis and more analytical
work. For the time being we point that
a drastic change of the weight for momenta differing
by the nesting wavevector $q_{\pi}=(\pi,\pi)$,  which are degenerate in
energy, is a remarkable prediction of
our approach which explains very well the numerical data  on
small systems and  can be
easily detected experimentally by photoemission experiments.
For instance
it  is not possible to obtain the above property within the Kane {\it et al.}
approach, because in this case the Green's function satisfies to
$G(k+q_{\pi},\omega)=G(k,\omega)$, so that the
weights $Z_k$ and $Z_{k+q_{\pi}}$ can only be equal.

Work is in progress  for extending the calculation at finite density
of holes and/or at smaller values of $J/t$.

S.S. acknowledges useful discussions with A. Parola and E.
Tosatti, and the kind hospitality at the University of Toulouse  and  at
the ITP in Santa Barbara.
A.A. acknowledges useful discussions with S. Klee and A. Muramatsu, and
financial support from the ``Human Capital and Mobility'' program under
contract $\#$ ERBCHBICT930475. D.P. acknowledges financial support from
the "Human Capital and Mobility" grant $\#$ CHRX-CT93-0332.
We thank IDRIS at Orsay (France) for allocation of CPU-time on the
Cray-C98 supercomputer.

\begin{table}
\caption{ Linear spin-wave estimates of the
 quasiparticle  weights and the one-hole energies
in the static limit ($t=0$) for the
various clusters. The percentage relative errors refer to the
spin-wave data as compared to the exact diagonalization results.}
\begin{tabular}{|c|c|c|c|c|}
\hline
M &  QP Weight  & \% Error & Energy & \%Error \\
\hline
8  & 1 &   0.0 &  2.5 & 0.0 \\
\hline
10 &  0.9919 & 0.0047 &  2.4308 & -0.35\\
\hline
16&             0.9724 & -0.22 &  2.3271 & -0.61 \\
\hline
18&             0.9688 & 0.055 & 2.3052 & -0.60 \\
\hline
20&             0.9637 & 0.050&  2.2908 & -0.65 \\
\hline
26&             0.9512 & 0.077 & 2.2609 & \\
\hline
32&             0.9422  & & 2.2415 & \\
\hline
$\infty$ &            0.820 & &   2.17 & \\
\hline
\end{tabular}
\end{table}

\begin{figure}
\caption{ (a)  Quasiparticle weight as a function of $t/J$.  The empty dots
correspond to our spin--wave results; the full dots are the exact results
on finite lattices. Continuous lines are guide to the eyes. (b) Same
notation as in (a) except that the empty dots refer to the $100$ sites
lattice. The dotted line is a local unstable minima for $t/J > 0.6$ }
\label{fig1}
\end{figure}

\unitlength1cm
\begin{picture}(15,7.0)
\put(0.9,-6.80){\includegraphics{fig1a.ps}}
\put(0.9,-16.20){\includegraphics{fig1b.ps}}
\end{picture}

\end{document}